\documentclass[10pt]{article}
\usepackage{amsfonts,amsmath,amssymb}
\usepackage{float}
\usepackage{pslatex}
\usepackage{subfigure}
\usepackage{graphicx,wrapfig}

\newcommand{\p}{\partial}

\newcommand{\mw}{\mathcal{W}}

\newcommand{\oB}{\overline{B}}
\newcommand{\occ}{\overline{c}}
\newcommand{\oG}{\overline{G}}
\newcommand{\atana}{\mbox{ArcTan}\left(\frac{1}{2}\frac{\sqrt{q^2}}{m}\right)}
\newcommand{\atanb}{\mbox{ArcTan}\left(\frac{\sqrt{q^2}}{m}\right)}
\newcommand{\woq}{\sqrt{q^2}}
\renewcommand{\d}{\ensuremath{\mathrm{d}}}
\usepackage[small,bf]{caption}
\begin{document}
\title{{\bf A gauge invariant infrared stabilization of $3D$ Yang-Mills gauge theories}}
\author{D.~Dudal\thanks{david.dudal@ugent.be}\,\,$^a$, J.~A.~Gracey\thanks{gracey@liv.ac.uk}\,\,$^b$, S.~P.~Sorella\thanks{sorella@uerj.br\;;\;Work supported by FAPERJ, Funda{\c
c}{\~a}o de Amparo {\`a} Pesquisa do Estado do Rio de Janeiro, under
the program {\it Cientista do Nosso Estado},
E-26/100.615/2007.}\,\,$^c$,
N.~Vandersickel\thanks{nele.vandersickel@ugent.be}\,\,$^a$ and
H.~Verschelde\thanks{henri.verschelde@ugent.be}\,\,$^a$\\\\
\small $^a$ Ghent University, Department of Mathematical Physics and
Astronomy \\ \small Krijgslaan 281-S9, 9000 Gent,
Belgium\\\\
$^b$ \small Theoretical Physics Division, Department of Mathematical Sciences, University of Liverpool\\ \small P.O. Box 147, Liverpool, L69 3BX, United Kingdom\\\\
$^c$ \small Departamento de F\'{\i }sica Te\'{o}rica, Instituto de
F\'{\i }sica, UERJ - Universidade do Estado do Rio de Janeiro\\
\small   Rua S\~{a}o Francisco Xavier 524, 20550-013 Maracan\~{a},
Rio de Janeiro, Brasil \normalsize}
\date{}
\maketitle\vspace{-0.5cm}
\begin{abstract}
\noindent We demonstrate that the inversion method can be a very
useful tool in providing an infrared stabilization of $3D$ gauge
theories, in combination with the mass operator $A^2$ in the Landau
gauge. The numerical results will be unambiguous, since the
corresponding theory is ultraviolet finite in dimensional
regularization, making a renormalization scale or scheme obsolete.
The proposed framework is argued to be gauge invariant, by showing
that the nonlocal gauge invariant operator $A^2_{\min}$, which
reduces to $A^2$ in the Landau gauge, could be treated in $3D$, in
the sense that it is power counting renormalizable in any gauge. As a corollary of our analysis, we are able to identify a whole set of powercounting renormalizable nonlocal operators of dimension two.
\end{abstract} 
\vspace{-14cm} \hfill LTH--782 \vspace{14cm}
\section{Introduction}
$3D$ gauge theories are not only of a pure theoretical importance.
For example, they naturally appear as the very high temperature
limit of their $4D$ counterpart \cite{Appelquist:1981vg}, while $3D$
QED can be used as an effective theory describing high temperature
cuprate superconductors \cite{qed3}.

When studying $3D$ gauge theories, a key observation is that the
gauge coupling $g^2$ becomes a dimensionful quantity. A positive
consequence is the ensuing superrenormalizability of $3D$ gauge
theories, meaning that the ultraviolet sector is relatively well
behaved. Unfortunately, this also inflicts a serious problem in the
infrared \cite{Jackiw:1980kv}. A simple dimensional counting allows
us to understand this problem intuitively. At consecutive orders in
perturbation theory in the massive coupling $g^2$, an increasing
number of inverse momenta is required to obtain a specific dimension
of e.g. a particular Green function under study. Consequently, at
increasing order of perturbation theory, the low momentum region
becomes more and more divergent.

A natural solution to this problem might be the dynamical generation
of a mass $m$, so that a perturbative expansion in the dimensionless
parameter $\frac{g^2}{m}$ might emerge, ensuring a safe infrared
limit. This has stimulated several studies,
\cite{Buchmuller:1994qy,Alexanian:1995rp,Jackiw:1995nf,Jackiw:1997jg,Karabali:1996iu,Heinz:1998eb}
to quote only a few. A common feature of the approaches of for
example
\cite{Buchmuller:1994qy,Alexanian:1995rp,Jackiw:1995nf,Jackiw:1997jg}
is that a dynamical gluon mass is derived from a certain gap
equation, constructed in a particular approximation scheme, bearing
nontrivial solutions.

The aim of this article is to investigate $3D$ gauge theories in the
presence of the gauge invariant nonlocal operator
\begin{equation}\label{A2min}
    A^2_{\min}=(VT)^{-1}\min_{U\in SU(N)}\int \d^3x
    \left(A_\mu^U\right)^2\,.
\end{equation}
In the first part of this paper, we shall prove its power counting renormalizability,
making the mass term $m^2A^2_{\min}$ a gauge invariant candidate for
an infrared regularization. As a nice byproduct of this proof, we shall be able to identify a whole class
of gauge invariant nonlocal operators which also enjoy the property of being UV powercounting renormalizable. However, we shall discuss why $A^2_{\min}$ plays a preferential role, since we must also take into account potential infrared problems. In the second part we shall employ the \emph{inversion method} to get a meaningful perturbative expansion for $3D$ gauge theories when the regulating mass coupled to $A^2_{\min}$ is brought back to
zero. This is done in the case of the Landau gauge, as $A^2_{\min}$
then reduces to the local operator $A^2$, which has already been
studied before, revealing several interesting properties
\cite{Dudal:2004ch,Dudal:2006ip}. We end with a discussion of the
results.

\section{The UV power counting renormalizability of $A^2_{\min}$ in $3D$}
\subsection{Preliminaries}
We intend to use the following action in $3D$ Euclidean space time
\begin{equation}\label{start}
S=\int \d^3x\left(\frac{1}{4}F_{\mu\nu}^2+b \p_\mu A_\mu +\occ
\p_\mu D_\mu c+\frac{1}{2}m^2A_\mu^2\right)\,,
\end{equation}
which corresponds to a Yang-Mills theory plus Landau gauge fixing,
supplemented with a regulating mass term. It was not only proven
that this action is renormalizable to all orders of perturbation
theory, but even that it is finite in dimensional regularization,
i.e. no renormalization is needed \cite{Dudal:2004ch,Dudal:2006ip}.

There are 2 remaining questions to be answered. The added mass term
$\propto A^2$ does not appear to be gauge invariant, and the
hitherto free mass parameter $m$ should be of a dynamical nature.
Since the $3D$ gauge coupling carries a dimension, one expects that
the dynamics of the theory will dictate  $m\propto g^2$.

The operator $A_\mu^2$ has a gauge invariant meaning in the Landau
gauge, $\p A=0$. Indeed, the gauge invariant operator \eqref{A2min}
can be rewritten as a perturbative series \cite{Lavelle:1995ty}
\begin{eqnarray}
A_{\min }^{2}
&=&\frac{1}{2}\int \d^{3}x\left[ A_{\mu }^{a}\left( \delta _{\mu \nu }-\frac{%
\partial _{\mu }\partial _{\nu }}{\partial ^{2}}\right) A_{\nu
}^{a}-gf^{abc}\left( \frac{\partial _{\nu }}{\partial ^{2}}\partial
A^{a}\right) \left( \frac{1}{\partial ^{2}}\partial {A}^{b}\right)
A_{\nu }^{c}\right] \;+O(A^{4})\,,  \label{min1}
\end{eqnarray}
and clearly $A^2_{\min}=A^2$ when $\p A=0$.

Recently, in $4D$, it has been argued that the highly nonlocal
operator $A^2_{\min}$ might be handled in a perturbative fashion
\cite{Gracey:2007ki}. The idea is to use the following termwise
gauge invariant representation of $A^2_{\min}$
\cite{Zwanziger:1990tn},
\begin{eqnarray}
A_{\min }^{2} &=&\mbox{Tr}\int \d^{3}x\left( F_{\mu \nu
}\frac{1}{D^{2}}F_{\mu \nu }+2ig\frac{1}{D^{2}}F_{\lambda \mu
}\left[ \frac{1}{D^{2}}D_{\kappa }F_{\kappa \lambda
},\frac{1}{D^{2}}D_{\nu }F_{\nu \mu }\right] \right.
\nonumber \\
&&-2ig\left. \frac{1}{D^{2}}F_{\lambda \mu }\left[
\frac{1}{D^{2}}D_{\kappa }F_{\kappa \nu },\frac{1}{D^{2}}D_{\nu
}F_{\lambda \mu }\right] \right) +O(F^{4})\,,  \label{serie1}
\end{eqnarray}
added to the action in the following form
\begin{eqnarray}
S_m&=&m^2\mbox{Tr}\int \d^{3}x\left(
F_{\mu \nu }\frac{1}{D^{2}}F_{\mu \nu }+2ig\frac{1}{D^{2}}F_{\lambda
\mu }\left[ \frac{1}{D^{2}}D_{\kappa }F_{\kappa \lambda
},\frac{1}{D^{2}}D_{\nu }F_{\nu \mu }\right] \right.
\nonumber \\
&&-2ig\left. \frac{1}{D^{2}}F_{\lambda \mu }\left[
\frac{1}{D^{2}}D_{\kappa }F_{\kappa \nu },\frac{1}{D^{2}}D_{\nu
}F_{\lambda \mu }\right] \right) +O(F^4)\,,  \label{serie2}
\end{eqnarray}
or in a more condensed notation
\begin{eqnarray}
S_m&=&m^2\mbox{Tr}\int \d^{3}x\left(\mathcal{O}_2+
g\mathcal{O}_3+g^2\mathcal{O}_4+\ldots\right)\,, \label{serie3}
\end{eqnarray}
where the $\mathcal{O}_k$ are gauge invariant functionals of the
field strength $F$ and covariant derivative $D$. As noticed in
\cite{Gracey:2007ki}, this expansion can be seen as one in operators
with $k$ legs where $k$ counts the lowest number of gluon legs
present in the operator $\mathcal{O}_k$.

Since the series \eqref{serie3} or \eqref{min1} contains infinitely
many nonlocal terms, it appears to be beyond our possibilities to
show that such a highly nonlocal operator might be renormalizable.
In general, the utmost we could achieve is to show the
renormalizability up to a certain (low) order. Therefore, we
consider the expansion \eqref{serie3}. Each of the nonlocal gauge
invariant terms could be studied separately. Such an approach was
successfully employed in $4D$ in the case of the first term,
$F\frac{1}{D^2}F$ \cite{Capri:2005dy,Capri:2006ne}. At the cost of
introducing a set of auxiliary bosonic and fermionic fields, the
action with the nonlocal operator added to it, was cast into a local
form. Using the many Ward identities of the resulting
action, we were able to prove the renormalizability to all orders of
perturbation theory. We started from
\begin{equation}\label{2} S=\int
\d^4x\left(\frac{1}{4}F_{\mu\nu}^aF_{\mu\nu}^a-\frac{1}{4}m^2F_{\mu
\nu }^{a}\left[ \left( D^{2}\right) ^{-1}\right] ^{ab}F_{\mu \nu
}^{b}\right)\,,
\end{equation}
recast into
\begin{eqnarray}\label{hen1}
  S_{local} &=& \int \d^4x\left[\frac{1}{4}F_{\mu \nu }^{a}F_{\mu \nu }^{a}+\frac{im}{4}(B-\overline{B})_{\mu\nu}^aF_{\mu\nu}^a
  +\frac{1}{4}\left( \overline{B}_{\mu \nu
}^{a}D_{\sigma }^{ab}D_{\sigma }^{bc}B_{\mu \nu
}^{c}-\overline{G}_{\mu \nu }^{a}D_{\sigma }^{ab}D_{\sigma
}^{bc}G_{\mu \nu }^{c}\right)\right]\,,
\end{eqnarray}
with $B,\oB$ a pair of complex bosonic antisymmetric tensor fields
in the adjoint representation and $G,\oG $ a pair of anticommuting
antisymmetric tensor fields, also in the adjoint representation. We
succeeded in constructing a gauge invariant classical action
$S_{cl}$ containing the mass parameter $m$ which is renormalizable.
This was proven to all orders of perturbation theory in the class of
linear covariant gauges, and explicitly checked up to two loop order
\cite{Capri:2005dy,Capri:2006ne}. In particular, this action reads
\begin{eqnarray}
  S_{phys} &=& S_{cl} +S_{gf}\;,\label{completeaction}\\
  S_{cl}&=&\int \d^4x\left[\frac{1}{4}F_{\mu \nu }^{a}F_{\mu \nu }^{a}+\frac{im}{4}(B-\overline{B})_{\mu\nu}^aF_{\mu\nu}^a
  +\frac{1}{4}\left( \overline{B}_{\mu \nu
}^{a}D_{\sigma }^{ab}D_{\sigma }^{bc}B_{\mu \nu
}^{c}-\overline{G}_{\mu \nu }^{a}D_{\sigma }^{ab}D_{\sigma
}^{bc}G_{\mu \nu
}^{c}\right)\right.\nonumber\\
&-&\left.\frac{3}{8}%
m^{2}\lambda _{1}\left( \overline{B}_{\mu \nu }^{a}B_{\mu \nu
}^{a}-\overline{G}_{\mu \nu }^{a}G_{\mu \nu }^{a}\right)
\right.\nonumber\\&+&\left.m^{2}\frac{\lambda _{3}}{32}\left(
\overline{B}_{\mu \nu }^{a}-B_{\mu \nu }^{a}\right) ^{2}+
\frac{\lambda^{abcd}}{16}\left( \overline{B}_{\mu\nu}^{a}B_{\mu\nu}^{b}-\overline{G}_{\mu\nu}^{a}G_{\mu\nu}^{b}%
\right)\left( \overline{B}_{\rho\sigma}^{c}B_{\rho\sigma}^{d}-\overline{G}_{\rho\sigma}^{c}G_{\rho\sigma}^{d}%
\right) \right]\,, \label{completeactionb}\\
S_{gf}&=&\int \d^{4}x\;\left( \frac{\alpha }{2}b^{a}b^{a}+b^{a}%
\partial _{\mu }A_{\mu }^{a}+\overline{c}^{a}\partial _{\mu }D_{\mu
}^{ab}c^{b}\right)\,.\label{lcg}
\end{eqnarray}
We draw attention to the fact that an additional quartic tensor
coupling $\lambda^{abcd}$, as well as two new mass couplings
$\lambda_1$ and $\lambda_3$ had to be introduced in order to
maintain renormalizability. This fact obscures the identification
with the original operator $F\frac{1}{D^2}F$, and hence with
$A^2_{\min}$. Nevertheless, at one loop, these extra couplings are
not yet relevant in the renormalization of $A^2_{\min}$, as found in
\cite{Gracey:2007ki}.  The renormalizability was explicitly
confirmed at 1-loop in a general linear covariant gauge, with a
gauge parameter independent anomalous dimension. The retrieved value
did coincide with the already known result for the anomalous
dimension of $A^2$ in the Landau gauge
\cite{Verschelde:2001ia,Gracey:2002yt}, as expected from the gauge
invariance of $A^2_{\min}$ and the perturbative equivalence with
$A^2$. In this sense, the result of \cite{Gracey:2007ki} is very
interesting as it provides evidence that $A^2_{\min}$ could be
consistently used at least at lowest order. Since it is gauge
invariant, one can opt to work in the Landau gauge, where it reduces
to a single local operator, which can enter the OPE for example.
Needless to say, this also considerably simplifies practical
calculations. However, so far, the analysis was restricted to lowest
order. Beyond the 1-loop approximation, little is known. Things
inevitably will become complicated since the new coupling
$\lambda^{abcd}$ will explicitly enter the analysis\footnote{The
2-loop anomalous dimension of $m$ is $\lambda^{abcd}$ dependent
\cite{Capri:2006ne}.}, and evidently, when $A^2_{\min}$ would be
renormalizable, its anomalous dimension is not supposed to contain
any new couplings.

In principle, a completely similar pathway could be followed in the
$3D$ case. One could investigate whether the consecutive terms in
the expansion \eqref{serie2} are renormalizable, by introducing
extra fields etc. Almost needless to say, this is still a very
cumbersome job.  For the first term $\mathcal{O}_2$, this would
amount to  a $3D$ analysis of its localized version similar to $4D$
\cite{Capri:2005dy,Capri:2006ne}. The Lorentz structure might be
simplified a bit by using the dual vector field
$f_\mu=\frac{1}{2}\varepsilon_{\mu\kappa\lambda}F_{\kappa\lambda}$
and analogs for the localizing fields. A potential caveat would be
the emergence of the extra couplings, which again make it obscure
(or even make it impossible) to say that $\mathcal{O}_2$ itself is a
renormalizable operator. However, it might be very well possible
that the massive $3D$ version of \eqref{completeaction} without
extra couplings ($\lambda_{1}=\lambda_{3}=\lambda^{abcd}=0$) is
renormalizable. We recall that these couplings were originally
introduced to absorb generated new counterterms. At 1-loop, the $3D$
theory ought to be finite, as the ``master integral'' is finite in
dimensional regularization \cite{Dudal:2006ip}, beyond 2-loops no
new counterterms can arise (similar arguments as in
\cite{Dudal:2006ip}), so the only possible source of these extra
couplings would lie at 2-loop order. In principle, this could be
checked similarly as done for $A^2$ in the Landau gauge
\cite{Dudal:2006ip}. In this work, we shall follow a slightly
different route, as the situation might be more appealing in $3D$
due to the superrenormalizability.

\subsection{Inductive proof of the UV power counting renormalizability of $A^2_{\min}$ }
In this section, we shall establish the UV power counting renormalizability of $A^2_{\min}$. As this proof will turn out to be rather technical, let us give a brief sketch of the main argument. We notice that in the expansion \eqref{serie1}, vertices will appear with an arbitrary power of the coupling $g$. Hence, as $g$ becomes dimensionful in $3D$, vertices with a certain power of $g$ will necessarily induce a certain power of compensating momenta in the denominator of the analytical expression corresponding to a Feynman diagram containing this vertex in order to obtain the correct dimensionality. Therefore, one may suspect that this $A^2_{\min}$ could be UV power counting renormalizable in $3D$.

Let us now put the previous line of intuitive reasoning on a more formal footing. We start the discussion from the operators $\mathcal{O}_k$ defined in expression \eqref{serie3}. These operators give rise to a set of new vertices.
We shall give a diagrammatical inductive argument that these vertices
are sufficiently suppressed in the UV such that \emph{no} new
ultraviolet divergences will appear.  The already present
counterterms\footnote{As far as these are existing of course, in
case the starting theory would be finite.} of the starting action
should be sufficient to render the complete theory finite.

The mass dimension of the operator $\mathcal{O}_k$ is actually given
by $\mbox{dim}\left[\mathcal{O}_k\right]=2-\frac{k}{2}$, since we
have $\mbox{dim}[g]=\frac{1}{2}$. We are in
$3D$, and each vertex is already multiplied by $m^2$. Consider a
vertex $V_i$ with $i$ ($i\geq2$) gluons legs present. We are
interested in the high momentum influence of such a vertex $V_{i}$,
i.e. we are interested in its UV ``cost'' for the renormalizability
analysis. Since $\mbox{dim}[A_\mu]=\frac{1}{2}$ and a vertex with
$i$ legs is multiplied by $g^{i-2}$, we can conclude that
$V_{i}\sim\frac{1}{Q^{i-2}}$, where we represent in general (a
combination of) momenta by the rather symbolic notation $Q$.

Consider now a random renormalizable set
of diagrams of the original theory. More precisely, we consider an
arbitrary set of (connected) $n$-point functions. Since the original
theory is supposed to be renormalizable, its $n$-point functions are
finite after including all the necessary counterterms. To begin, we
want to attach a \emph{single} vertex $V_i$ onto these $n$-point
functions, in order to obtain a certain $n'$-point function. One can check that we obtain all possibilities just
by connecting a series of external gluon legs. As a consequence
there are only two possible operations we can undertake:
\begin{enumerate}
\item We can attach an external gluon leg of the vertex $V_i$ to an external gluon leg of an arbitrary $n$-point function. Take $\ell$ the number of gluon legs of $V_i$ attached
to original diagrams, then $1 \leq \ell \leq i$.

\item We can also glue two gluon legs of the vertex $V_i$ together. We assume that in total $2s$ legs are pairwise closed on each other to form loops.
\end{enumerate}
After carrying out these operations, $e$ gluon legs of the vertex $V_i$ remain and will serve as external legs. Hence, we have
\begin{eqnarray}\label{som}
2s + \ell + e &=& i\,.
\end{eqnarray}
We notice that the above procedure will generate \emph{all} possible
$n'$-point functions of the new theory in which a single new vertex
$V_i$ has been used, as the original starting set was arbitrary.

We shall now search for an estimate of the UV ``weight''
$\mathcal{W}$ caused by the new vertex insertion. We shall work in a
``worst case scenario'', i.e. we always consider the case that the
UV behaviour is the worst of possible occurring scenarios. In order to make things as comprehensible as possible, we shall explore all possible occurring scenarios one by one.

\subsubsection{The vertex}
As already explained before, the vertex $V_i$ will induce a weight
\begin{eqnarray}
  \mw_V &\sim& \frac{1}{Q^{i-2}}\,.
\end{eqnarray}
\subsubsection{The $s$ loops}
As it is easily checked, the $2s$ legs closing on each other will generate the following contribution after integration
\begin{eqnarray}
  \mw_s &\sim& \int \frac{(\d^3Q)^s}{(Q^{2})^s}\sim Q^s\,.
\end{eqnarray}
\subsubsection{The attachment of the $\ell$ legs}
Each of the $\ell$ legs of $V_i$ shall be glued to an external leg
of an original diagram, which is part of a renormalizable set. Such
an external leg is coming from a $3$- or $4$-point
vertex\footnote{The $3$ refers to the 3-gluon and ghost-gluon
vertex. For simplicity, we have omitted the quarks for the moment.}
of the original Yang-Mills theory. We will consider case by case,
whereby every case is determined by the number of legs of $V_i$ that
``arrives'' at the same vertex of the original diagram.
\begin{itemize}
  \item \textbf{Case 1: the single attachment}\\
  Let us assume that there are $\ell_1$ spots at which a single leg arrives. There are 4 scenarios, as shown in Figure 1:
  \begin{enumerate}
    \item \emph{$\ell_{1a}'$ by using a 3-vertex with 2 external legs}\\
    As clearly depicted in Figure \ref{fig1a}, this corresponds to using $\ell_{1a}'$ times a tree level 3-point vertex as starting point. Each time, one leg of it is glued to a leg of the new vertex $V_i$. The UV weight is obtained as
     \begin{eqnarray}
       \mw_{\ell_{1a}'} &\sim& \left(\frac{1}{Q^2}Q\right)^{\ell_{1a}'}\sim
       \frac{1}{Q^{\ell_{1a}'}}\,.
     \end{eqnarray}
     The $\frac{1}{Q^2}$ corresponds to the extra propagator caused by gluing 2 legs together, the $Q$ comes from the 3-point vertex\footnote{Notice that this is a worst case scenario, as the ghost-gluon vertex does not carry a momentum factor.}, and there are no loop integrations possible in this case.

    \item \emph{$\ell_{1b}'$ by using a 3-vertex with 1 or 0 external legs}\\
    In comparison with the previous case, one of the legs no longer serves as an external leg, but becomes connected itself to another vertex of the old diagram.
This possibility is depicted in Figure \ref{fig1b}, where the grey
area stands for any other diagram\footnote{In order not to overload
the picture, we did not draw (the) other leg(s) connected to the
grey blob assuring that the final diagram would be 1PI.}. In the
current case, one can create additional loops which will influence
the to-be-derived weight factor. In addition to the foregoing weight
factor, we shall also encounter extra loop integrals. In particular,
     \begin{eqnarray}\label{slechteweight}
       \mw_{\ell_{1b}'} &\sim& \left(\int \d^3Q\frac{1}{Q^2}Q\right)^{\ell_{1b}'}\sim
       (Q^2)^{\ell_{1b}'}\,.
     \end{eqnarray}
     However, this counting procedure is not as fine as desired to obtain a reasonable final result, being a sufficiently suppressed UV weight. Fortunately, there is an
alternative way of obtaining such kind of vertex insertion, which
allows for a refined weight factor. Namely, taking a look at Figure
\ref{fig1b}, we see that the original diagram can be thought of as
another diagram of the old theory that we have cut open, added an
extra old 3-vertex to it, and then we have attached to this extra
vertex the legs coming from the new vertex $V_i$. The upshot of this
viewpoint is that now, we can take into account that the ``cut \&
paste'' operation on the new original diagram brings an extra
propagator, viz. $\frac{1}{Q^2}$, into the game. As such, we obtain
     \begin{eqnarray}\label{goedeweight}
       \mw_{\ell_{1b}'} &\sim& \left(\int \d^3Q\frac{1}{(Q^2)^2}Q\right)^{\ell_{1b}'}\sim
       (1)^{\ell_{1b}'}=1
     \end{eqnarray}
     rather than \eqref{slechteweight}.\\\\
     A word of caution on the emerging loop integrals: it should be understood that these loop integrations are performed
at the end, when all attaching operations are done. However, for the
purpose of counting the UV weight, we have distributed them over the
several subcases, since this does not influence the formal power
counting and allows for a more efficient bookkeeping.

     \item \emph{$\ell_{1a}''$ by using a 4-vertex with 3 external legs}\\
     This case (see Figure \ref{fig1c}) can be treated in an analogous fashion as the $\ell_{1a}'$ case, albeit no $Q$ will appear since  a 4-vertex does not contain momentum
dependent factors. We find
     \begin{eqnarray}
       \mw_{\ell_{1a}''} &\sim& \frac{1}{(Q^2)^{\ell_{1a}''}}\,.
     \end{eqnarray}

     \item \emph{$\ell_{1b}''$ by using a 4-vertex with 2,1 or 0 external legs}\\
     This is analogous as the $\ell_{1b}'$ case, but without the $Q$, thus yielding
     \begin{eqnarray}
       \mw_{\ell_{1b}''} &\sim& \left(\int \d^3Q\frac{1}{(Q^2)^2}\right)^{\ell_{1b}''}\sim
       \frac{1}{Q^{\ell_{1b}''}}\,.
     \end{eqnarray}
     The corresponding diagram is shown in Figure \ref{fig1d}.
     \end{enumerate}

  \begin{figure}[t]
  \begin{center}
    \subfigure[]{\includegraphics[width=2.5cm]{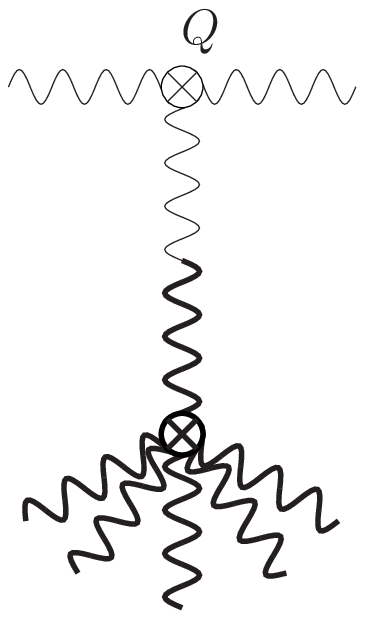} \label{fig1a}}
    \hspace{1cm}
    \subfigure[]{\includegraphics[width=2.5cm]{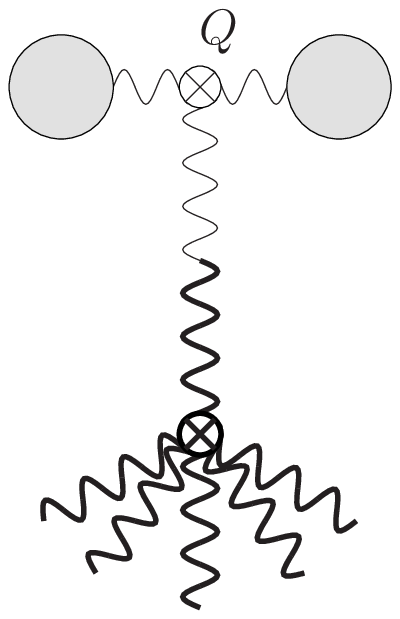}\label{fig1b}}
    \hspace{1cm}
    \subfigure[]{\includegraphics[width=2.5cm]{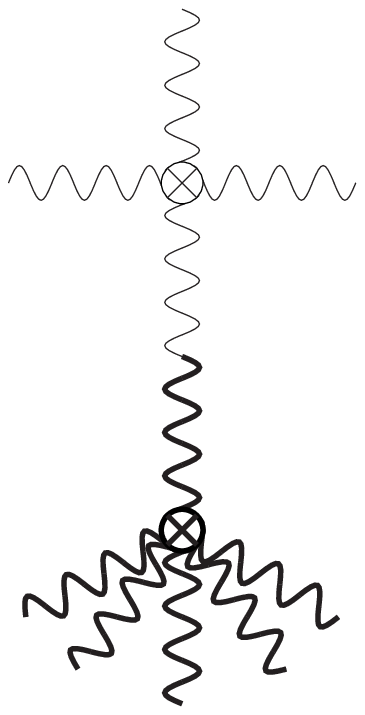}\label{fig1c}}
    \hspace{1cm}
    \subfigure[]{\includegraphics[width=2.5cm]{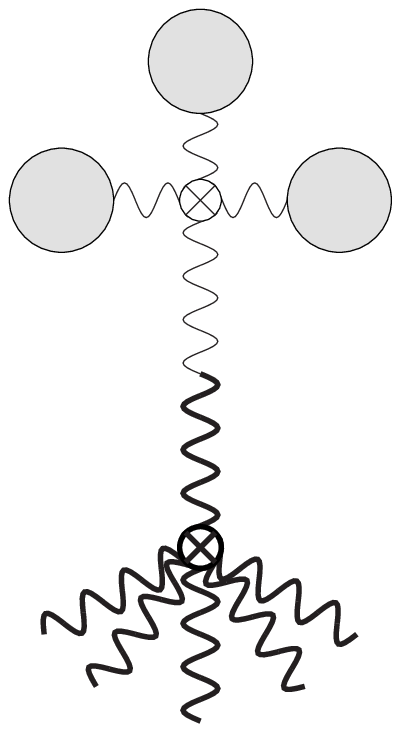}\label{fig1d}}
  \end{center}
  \caption{The possible Feynman diagram configurations for Case 1. Fat lines refer to the new vertex $V_i$, while normal lines to the original theory.}
\end{figure}

  \item \textbf{Case 2: the double attachment}\\
  Let us assume that there are $\ell_2$ spots at which 2 legs arrive (see Figure 2), whereby evidently each time a loop is created.   Also here,
  4 configurations
  arise:
  \begin{enumerate}
    \item \emph{$\ell_{2a}'$ by using a 3-vertex with 1 external leg}\\
    The following UV weight is found
     \begin{eqnarray}
       \mw_{\ell_{2a}'} &\sim& \left(\int \d^3Q\frac{1}{(Q^2)^2}Q\right)^{\ell_{2a}'}\sim
       1\,.
     \end{eqnarray}
For the benefit of the reader, let us again explain the origin of
the different components in the previous weight factor. There is a
loop integral, the two propagators building the loop and an extra
$Q$ corresponding to the 3-vertex (cfr Figure \ref{fig2a}).

    \item \emph{$\ell_{2b}'$ by using a 3-vertex with 0 external legs}\\
    Using a slightly adapted argument, namely a pure ``paste'' one, we obtain in this case (see Figure \ref{fig2b})
    \begin{eqnarray}
       \mw_{\ell_{2b}'} &\sim& \left(\int (\d^3Q)^2 \frac{1}{(Q^2)^3}Q\right)^{\ell_{2b}'}\sim
       Q^{\ell_{2b}'}\,.
     \end{eqnarray}
     Depicting the procedure for the reader: we attached 2 legs of the new vertex $V_i$ to a single external leg of the grey blob by means of 3-gluon vertex.

    \item \emph{$\ell_{2a}''$ by using a 4-vertex with 2 external legs}\\
    Now one recovers (see Figure \ref{fig2c})
    \begin{eqnarray}
       \mw_{\ell_{2a}''} &\sim& \left(\int \d^3Q \frac{1}{(Q^2)^2}\right)^{\ell_{2a}''}\sim
       \frac{1}{Q^{\ell_{2a}''}}\,.
     \end{eqnarray}

    \item \emph{$\ell_{2b}''$ by using a 4-vertex with 1 or 0 external legs}\\
    For the fourth double attachment scenario (see Figure
    \ref{fig2d}), we have
    \begin{eqnarray}
       \mw_{\ell_{2b}''} &\sim& \left(\int (\d^3Q)^2 \frac{1}{(Q^2)^3}\right)^{\ell_{2b}''}\sim
       1\,,
     \end{eqnarray}
     which is obtained by the ``cut \& paste'' logic.

  \end{enumerate}

  \begin{figure}[t]
  \begin{center}
    \subfigure[]{\includegraphics[width=2.5cm]{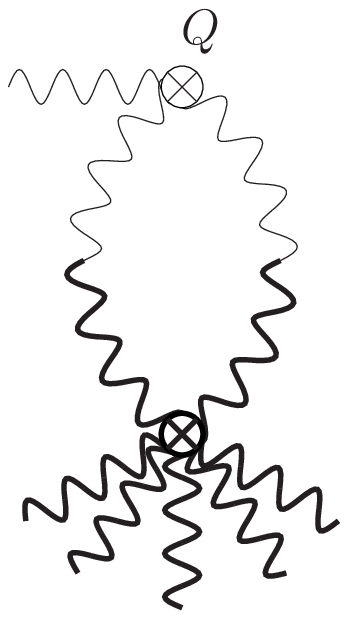}\label{fig2a}}
    \hspace{1cm}
    \subfigure[]{\includegraphics[width=2.5cm]{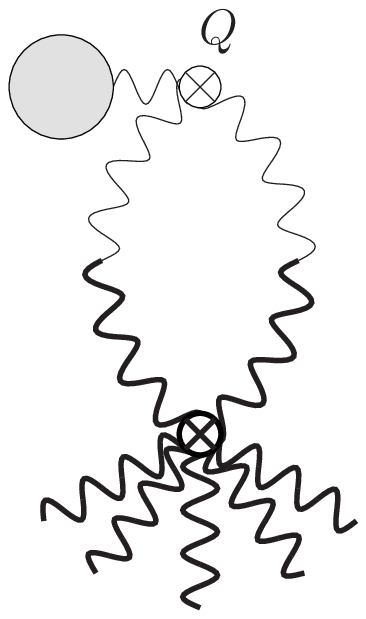}\label{fig2b}}
    \hspace{1cm}
    \subfigure[]{\includegraphics[width=2.5cm]{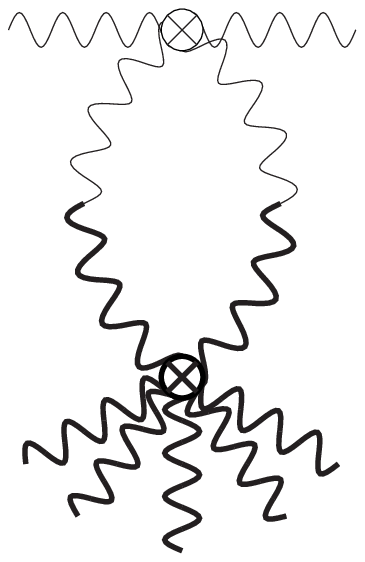}\label{fig2c}}
    \hspace{1cm}
    \subfigure[]{\includegraphics[width=2.5cm]{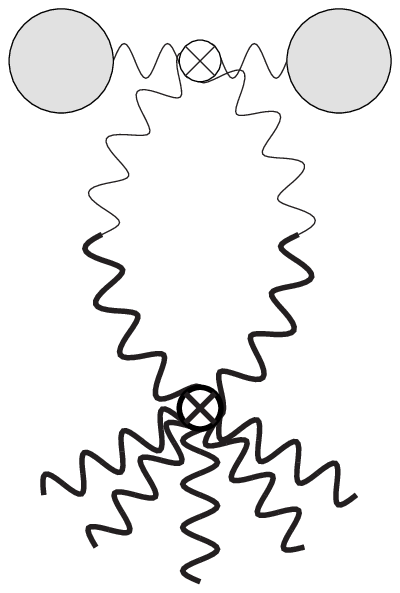}\label{fig2d}}
  \end{center}
  \label{fig2}
  \caption{The possible Feynman diagram configurations for Case 2.}
\end{figure}

  \item \textbf{Case 3: the triple attachment}\\
  Let us assume that there are $\ell_3$ spots at which 3 legs arrive. In this case, we observe 3 options:
  \begin{enumerate}
    \item \emph{$\ell_{3}'$ by using a 3-vertex}\\
    This is the simplest case, as no external legs are available. We find
    \begin{eqnarray}
       \mw_{\ell_{3}'} &\sim& \left(\int (\d^3Q)^2 \frac{1}{(Q^2)^3}Q\right)^{\ell_{3}'}\sim
       Q^{\ell_{3}'}\,.
     \end{eqnarray}
    There is a double loop integral, as it immediately follows from the diagram displayed in Figure \ref{fig3a}

    \item \emph{$\ell_{3a}''$ by using a 4-vertex with 1 external leg}\\
    From Figure \ref{fig3b}, we find
    \begin{eqnarray}
       \mw_{\ell_{3a}''} &\sim& \left(\int (\d^3Q)^2 \frac{1}{(Q^2)^3}\right)^{\ell_{3a}''}\sim
       1\,.
     \end{eqnarray}

    \item \emph{$\ell_{3b}''$ by using a 4-vertex with 0 external legs}\\
    Once more employing the ``cut \& paste'' argument leads us to
    \begin{eqnarray}
       \mw_{\ell_{3b}''} &\sim& \left(\int (\d^3Q)^3 \frac{1}{(Q^2)^4}\right)^{\ell_{3b}''}\sim
       Q^{\ell_{3b}''}\,.
     \end{eqnarray}
    The corresponding diagram is shown in Figure \ref{fig3c}.

  \end{enumerate}
  \begin{figure}[t]
  \vspace{1cm}
  \begin{center}
    \subfigure[]{\includegraphics[width=2.5cm]{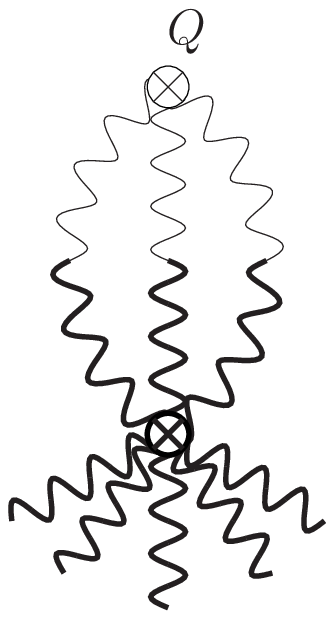}\label{fig3a}}
    \hspace{1cm}
    \subfigure[]{\includegraphics[width=2.5cm]{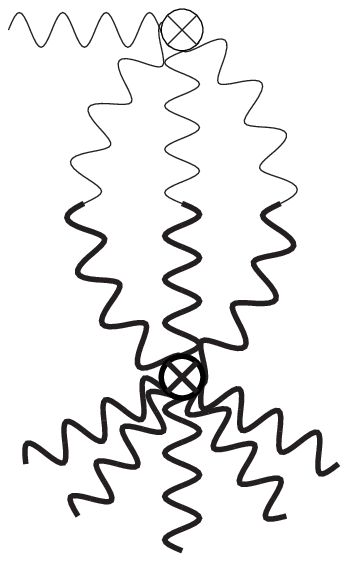}\label{fig3b}}
    \hspace{1cm}
    \subfigure[]{\includegraphics[width=2.5cm]{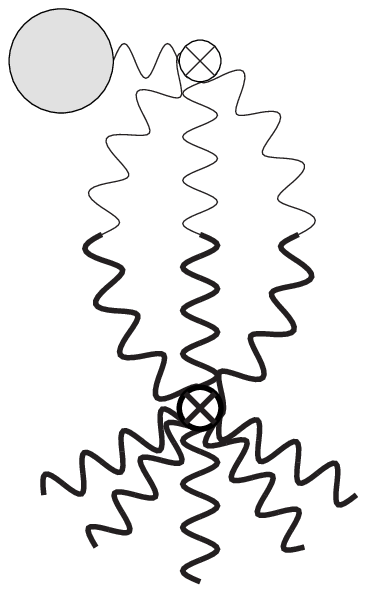}\label{fig3c}}
  \end{center}
  \caption{The possible Feynman diagram configurations for Case 3.}
\end{figure}

  \item \textbf{Case 4: the quadruple attachment}\\
  Finally, we can assume that there are $\ell_4$ spots at which 4 legs arrive. Evidently, only one possibility pops up:
    \begin{eqnarray}
       \mw_{\ell_{4}} &\sim& \left(\int (\d^3Q)^3 \frac{1}{(Q^2)^4}\right)^{\ell_{4}}\sim
       Q^{\ell_{4}}\,.
     \end{eqnarray}
     This was obtained analogously as in Case 2.2.
      \begin{figure}[t]
  \begin{center}
  \vspace{0.6cm}
    \includegraphics[width=2.5cm]{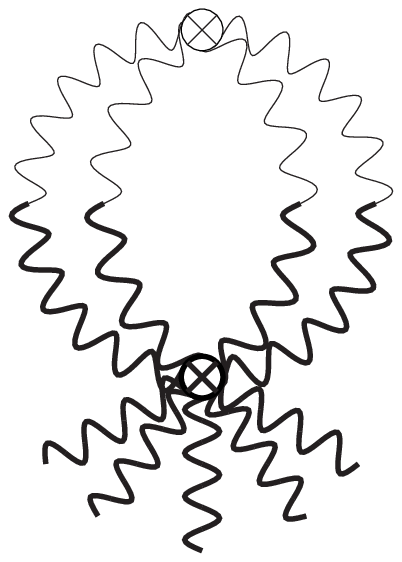}
  \end{center}
  \label{fig4}
  \caption{The possible Feynman diagram configuration for Case 4.}
\end{figure}
\end{itemize}
For later use, we mention that
\begin{eqnarray}\label{ell}
\ell&=&
\underbrace{\ell_{1a}'+\ell_{1b}'+\ell_{1a}''+\ell_{1b}''}_{=\ell_1}+2\underbrace{(\ell_{2a}'+\ell_{2b}'+\ell_{2a}''+\ell_{2b}'')}_{=\ell_2}+3\underbrace{(\ell_{3}'+\ell_{3a}''+\ell_{3b}'')}_{=\ell_3}+4\ell_4\,,
\end{eqnarray}
which is readily checked.

We are now ready to combine all the obtained information into a
single estimate for the UV weight,
\begin{eqnarray}\label{gewicht1}
       \mw &\sim& \frac{1}{Q^3} \mw_V\mw_s\mw_{\ell_{1a}'}\mw_{\ell_{1b}'}\mw_{\ell_{1a}''}\mw_{\ell_{1b}''}\mw_{\ell_{2a}'}\mw_{\ell_{2b}'}
       \mw_{\ell_{2a}''}\mw_{\ell_{2b}''}\mw_{\ell_{3a}'}\mw_{\ell_{3a}''}\mw_{\ell_{3b}''}\mw_{\ell_{4}}\,.
\end{eqnarray}
We introduced a factor $\frac{1}{Q^3}$, which serves as a ``correcting'' weight factor related to local momentum conservation at the vertex $V_i$. During the derivation of the different weight factors, we have always assumed that the introduced loop integrals were independent, but momentum conservation at $V_i$ will at least kill one of these. Let us also mention here that the $e$ external momenta of $V_i$ will get related to the other external momenta of the final diagram by means of global momentum conservation. \\
Simplifying expression \eqref{gewicht1}, we find for the total weight:
\begin{eqnarray}\label{gewicht2}
       &&\mw ~\sim~ \frac{1}{Q^{\kappa}} \;, \\
       &&\text{with} \quad \kappa = (i - 2) + 3 - s +  \ell_{1a}' + 2 \ell_{1a}'' + \ell_{1b}'' - \ell_{2b}' + \ell_{2a}'' - \ell_3' - \ell_{3b}'' -
       \ell_4\,.\nonumber
\end{eqnarray}
The power $\kappa$ can be further simplified by means of \eqref{som} and \eqref{ell}, which leads to
\begin{eqnarray}\label{gewicht3}
\kappa = 1 + s + e +2\ell_{1a}' + \ell_{1b}' + 3\ell_{1a}''
+2\ell_{1b}'' + 2\ell_{2a}'+  \ell_{2b}' + 3\ell_{2a}''+2\ell_{2b}''
+2 \ell_{3}''+3\ell_{3a}''+2\ell_{3b}''+3\ell_4>0\,.
\end{eqnarray}
Recapitulating, the power of $\frac{1}{Q}$ is strictly positive,
meaning that the UV weight $\mw$, \eqref{gewicht2}, of the new
vertex is under control very well when $Q\sim\infty$.

So far, we have ignored the presence of quarks. The counting
analysis will however remain valid even in that case. The only cases
in which the 3-point quark-gluon vertex will appear and influence
the power counting, correspond to the diagrams displayed in the
figures \ref{fig1a} or \ref{fig1b}. However, the
situation corresponding to the case \ref{fig1a} is
even further improved since the quark-gluon vertex does not contain
an explicit momentum dependence. For the case of \ref{fig1b}, we
notice that the extra quark propagator will behave as $\frac{1}{Q}$
instead of $\frac{1}{Q^2}$, but since there is no $Q$ coming from
the vertex itself, the eventual counting remains unaltered since
``$\frac{1}{Q^2}Q=\frac{1}{Q}$''.

To close the inductive argument, we can of course repeat the
previous argument when we allow a second, third, $\ldots$ new
vertex of the type $V_i$ into where we are now: the theory plus one
single vertex. The latter one has just been shown to behave well in
the UV. We conclude that any number of new vertices added to the
theory will only induce UV harmless additional (pieces of) diagrams.

Before closing this section, a few more words are to be devoted to the $i=2$ case. Setting $i=2$ (and thus $s=e=0$) in
the UV estimate \eqref{gewicht2}, we see that also the 2-legged
insertion is UV safe when used to
construct new diagrams starting from an originally
renormalizable theory. In this case, this would be massless $3D$ YM,
whose perturbation theory is unfortunately ill-defined. We can
however circumvent this problem. In every ``conventional'' gauge,
there exists a renormalizable (gauge variant) mass operator: the
Landau \cite{Dudal:2002pq}, the linear covariant
\cite{Dudal:2003np}, the Curci-Ferrari \cite{Dudal:2003pe}, the
maximal Abelian \cite{Dudal:2003pe} and a class of nonlinear
covariant gauges \cite{Lemes:2006aw}. Strictly speaking, this was
proven only in $4D$, with the exception of the Landau and
Curci-Ferrari gauges which were also explicitly analyzed in $3D$
\cite{Dudal:2004ch}, but the algebraic renormalizability analysis of
the mass operators does not really depend on the space time
dimension, and all the relevant Ward identities will remain valid in
$3D$. Continuing our reasoning, we choose a gauge to work in and add
an IR regulating mass term to it. Starting from this theory, we can
apply the foregoing arguments of this subsection also for the
2-legged insertion, and consequently conclude that $A^2_{\min}$ is
power counting renormalizable in any of these gauges. Once this is
established, we can drop again the temporarily introduced gauge
variant mass term and immediately add the gauge invariant mass term
$\propto A^2_{\min}$ to the action.

Summarizing, we have thus demonstrated that $A^2_{\min}$ should be
power counting renormalizable in $3D$. The gauge invariant mass
operator $A_{\min}^2$ should thus be consistent with the ultraviolet
renormalizability. For practical calculations, it would still be technically challenging to calculate with $A^2_{\min}$ even in the
Feynman gauge. However, since it is explicitly gauge invariant, we
do not make any sacrifices choosing to work in the clearly
preferable Landau gauge, in which case we have the relatively simple
action \eqref{start} at our disposal. In this case, we also do not
have to worry about the overabundance of $\frac{1}{\p^2}$ in the
expressions \eqref{min1} or \eqref{serie1}, which could cause IR
troubles during the calculation of the Feynman integrals. Due to the
gauge invariance and the subsequent choice of the Landau gauge where
no potentially dangerous $\frac{1}{\p^2}$ terms are present, it is
obvious that these possible infrared divergences cannot influence
physical quantities (they should cancel out if occurring in
intermediate results). We shall come back to this issue at the end
of section 3.

\subsection{A more general class of powercounting renormalizable gauge invariant 2D operators and the preferred role of $A^2_{\min}$}
 As the attentive reader might have noticed, we can extract more interesting information from our proof than simply the UV powercounting renomalizability of $A^2_{\min}$. In fact, all the operators $\mathcal{O}_i$ are separately UV powercounting renomalizable. A fortiori, so is any (infinite) linear combination of those. However, there is more to the story in $3D$ than just UV powercounting renormalizability, we should also take the infrared safety into account. First of all, $\mathcal{O}_2$ should be present as this is the only one that will give rise to a mass in the gluon propagator, which can serve as a natural infrared cut-off. Secondly, we should also be aware of the potential infrared danger caused by the $\frac{1}{\p^2}$'s in the interaction terms. It is exactly our point that by using a particular series of these operator, viz. $A^2_{\min}$, one can motivate that no infrared dangerous terms will occur when calculating gauge invariant quantities, as we have done at the end of the previous subsection.

\section{Removal of the regulating mass parameter $m$}
\subsection{The inversion method}
Albeit that we have regularized the $3D$ gauge theory in a gauge
invariant fashion by the introduction of the mass $m$, it is still a
mass introduced by hand. The next goal is to get rid of this
arbitrary parameter.

We shall explain the inversion method with the example of the pole
mass.  Consider the one loop gauge boson polarization tensor
$\pi_{\mu\nu}^{ab}$, which can be decomposed in the traditional way
\begin{equation}\label{1}
    \pi_{\mu\nu}^{ab}(q^2)=\delta^{ab}\left(\delta_{\mu\nu}-\frac{q_\mu
    q_\nu}{q^2}\right)\pi(q^2)+\delta^{ab}\frac{q_\mu
    q_\nu}{q^2}\omega(q^2)\,.
\end{equation}
It is then easily shown that the corresponding shift in the tree
level mass will be given by
\begin{equation}\label{2}
    m^2\to m^2+\pi(q^2)\,.
\end{equation}
The pole mass (squared) $m_p^2$ is defined as that value of $(-q^2)$
such that $ q^2+m^2+\pi(q^2)=0$. As we are working with a
perturbative series, this can be solved in an iterative way, so that
at lowest order
\begin{equation}\label{3}
    m_p^2= m^2+g^2\pi_1(-m^2)\,,
\end{equation}
where $\pi_1(q^2)$ is the one loop contribution to the self energy.

Having found an expression for the pole mass $m_p$ in terms of the
regulating mass $m$,
\begin{equation}\label{5}
    m_p^2=m^2\left(1+a_1 \frac{g^2}{m}+a_2\frac{g^4}{m^2}+\ldots\right)\equiv
    m^2\mathcal{A}(m^2)\,,
\end{equation}
the question remains what we must do with it, since $m$ should in
principle become zero again at the end of any calculation, to
restore equivalence with the original starting \emph{massless} YM
action. One option is to simply set $m=0$ in \eqref{5}. However, as
outlined in \cite{Fukuda:1988iv}, (certain) nonperturbative effects
can be taken into account when the series \eqref{5} is inverted as
follows,
\begin{equation}\label{6}
    m^2=m_p^2\left(1+b_1 \frac{g^2}{m_p}+b_2\frac{g^4}{m_p^2}+\ldots\right)\equiv
    m_p^2\mathcal{B}(m_p^2)\,.
\end{equation}
Apparently, $m=0$ can now be realized in 2 ways: by setting $m_p=0$
corresponding to the perturbative (potentially ill-defined)
solution, but also by solving the gap equation
\begin{equation}\label{7}
    \mathcal{B}(m_p^2)=0\,,
\end{equation}
which can give rise to a nontrivial solution $m_p^2\neq0$, whereby
that nevertheless $m=0$!

We used the example of the pole mass $m_p$ to explain the inversion
philosophy, but also other quantities ${\cal Q}$ could be handled:
the regulating mass $m$ is introduced to ensure a meaningful
perturbative series for ${\cal Q}(m)$, and after inversion a
meaningful (finite) result can be found for ${\cal Q}$ even for
$m=0$, by solving the gap equation $m({\cal Q})=0$. Hence, it
appears that the inversion method can be a very useful tool to
obtain results in superrenormalizable quantum field theories, which
are plagued by infrared instabilities.

\subsection{Explicit calculations}
We now turn to an explicit computation. We have calculated the gauge
boson self energy at one loop by evaluating the contributing four
Feynman diagrams in three dimensions where we also include massless
quarks. The diagrams are generated by the {\sc Qgraf} package,
\cite{Nogueira:1991ex}, and converted to {\sc Form} input notation
where {\sc Form} is a symbolic manipulation language,
\cite{Vermaseren:2000nd}. The self energy is then reduced to a set
of master one loop integrals which have previously been determined
in \cite{Rajantie:1996cw}. Whilst these have been deduced in
dimensional regularization in $d$~$=$~$3$~$-$~$2\epsilon$, the
relevant one loop integrals are in fact finite in three dimensions.
Hence we find that the transverse part of the self energy is
\begin{eqnarray}\label{pi1}
\pi(q^2)&=&-\frac{g^2}{8}T_F
N_f\sqrt{q^2}+\frac{g^2N}{\pi}\left\{\frac{7}{32}m-\frac{1}{32}\frac{m^3}{q^2}+\frac{5}{32}\frac{q^2}{m}+\frac{5}{8}\woq\atana\right.\nonumber\\
&&\left.-\frac{1}{2}\woq\frac{m^2}{q^2}\atana+\frac{1}{8}\woq\frac{q^2}{m^2}\atana\right.\nonumber\\
&&\left.-\frac{1}{64}\woq\frac{(q^2)^2}{m^4}\atana-\frac{5}{16}\woq\atanb\right.\nonumber\\
&&\left.+\frac{1}{32}\woq\frac{m^4}{(q^2)^2}\atanb-\frac{1}{8}\woq\frac{m^2}{q^2}\atanb\right.\nonumber\\
&&\left.
-\frac{1}{8}\woq\frac{q^2}{m^2}\atanb+\frac{1}{32}\woq\frac{(q^2)^2}{m^4}\atanb\right\}\nonumber\\
&&+g^2N\left\{\frac{1}{64}\woq-\frac{1}{128}\woq\frac{(q^2)^2}{m^4}\right\}
\end{eqnarray}
for the transverse component, and
\begin{eqnarray}
\omega(q^2)&=&
\frac{g^2N}{\pi}\left\{-\frac{1}{16}m+\frac{1}{16}\frac{m^3}{q^2}-\frac{1}{16}\woq\atanb-\frac{1}{16}\woq\frac{m^4}{(q^2)^2}\atanb\right.\nonumber\\
&&\left.-\frac{1}{8}\woq\frac{m^2}{q^2}\atanb\right\}+g^2N\frac{\woq}{32}
\end{eqnarray}
for the longitudinal component. The reader might be a little confused, as the computed 1-loop self
energy $\pi_{\mu\nu}^{ab}$ is apparently not transverse. However,
the Ward identity in the massive Landau case does not predict a
transverse self energy\footnote{Evidently, the propagator (connected
2-point function) is still transverse.}. Details can be found in the Appendix.

\subsection{Back to the pole mass} Let us return to the determination of the pole
mass, at 1-loop given by
\begin{equation}
    m_p^2=m^2+g^2\pi_1(-m^2)\,.
\end{equation}
A few complications arise. A first one is the appearance
of\footnote{We recall that
$\mbox{ArcTan}(z)=\frac{1}{2i}\ln\frac{1+iz}{1-iz}$.}
$\mbox{ArcTan}(i)=i\infty$ in $\pi(-m^2)$, but fortunately, these
terms cancel amongst each other\footnote{As a matter of fact, also
the longitudinal part is finite at $q^2=-m^2$.}. Less fortunately,
the presence of $\sqrt{q^2}$ results in a complex valued pole mass.
We find
\begin{equation}\label{poolqcd}
m_p^2=m^2-\frac{i}{8}T_F N_f
g^2m+\frac{N}{\pi}\left(\frac{3}{32}-\frac{63}{128}\ln3\right)g^2m+iN\frac{1}{128}g^2m+\ldots\,,
\end{equation}
since $\mbox{ArcTan}\left(\frac{i}{2}\right)=\frac{i}{2}\ln 3$.
Performing the inversion and solving the induced gap equation, we
obtain
\begin{equation}\label{poolqcd2}
m_p=-\frac{i}{8}T_F N_f
g^2+\frac{N}{\pi}\left(\frac{3}{32}-\frac{63}{128}\ln3\right)g^2+iN\frac{1}{128}g^2+\ldots\,.
\end{equation}
It is interesting that, using a different approach and different
mass operators, the works \cite{Jackiw:1995nf,Jackiw:1997jg} also
report a complex pole mass in some cases, depending on the employed
(gauge invariant) mass operator.

We recall here that $3D$ gauge theories are also confining. This
means that the gluon itself is not a physical particle, and as such,
unitarity should not be expected at the level of the elementary
gluon excitations. Therefore, we certainly do not claim that we have
obtained massive gauge bosons with 3 physical polarizations, even if
the eventual pole mass would have been real-valued. Notice that
unitarity at the level of the gluons was also not mentioned in e.g.
\cite{Jackiw:1995nf,Jackiw:1997jg}. The main point of these papers
and the current work is to find a way to regulate the $3D$ gauge
theory to allow for a consistent expansion. Once this is done, one
could try to have a look at the physical excitations, which are
supposed to be massive glueball states, a fact supported by the $3D$
lattice data \cite{Teper:1998te}. For example, one could try to
construct a bound state of gluons using standard techniques, or one
could study gauge invariant correlators like $\langle
F_{\mu\nu}^2(x) F_{\mu\nu}^2(y)\rangle$, since the operator
$F_{\mu\nu}^2$ has the correct quantum numbers to create/annihilate
a scalar glueball. This is however beyond the scope of this paper,
but let us only underline that an explicit calculation of $\langle
F_{\mu\nu}^2(x) F_{\mu\nu}^2(y)\rangle$ would also be plagued by
infrared singularities in $3D$, unless some regulating mechanism is
being provided.

\subsection{Applying the inversion method to the gluon propagator}
As a second example, we shall now determine an estimate for the
gluon propagator. In our conventions, the gluon propagator reads
\begin{equation}\label{prop2}
    D(p)=\frac{1}{p^2+m^2+\pi(p^2,m^2)}\,,
\end{equation}
which we have calculated explicitly to 1-loop order. For any value
of the Euclidean momentum $p=p_\ast$, we can invert \eqref{prop2} to
get $m^2=\mathcal{F}(D(p_\ast))$, and then solve
$\mathcal{F}(D(p_\ast))=0$ in order to obtain a value for
$D(p_\ast)$. Since we have
\begin{equation}\label{prop3}
    \frac{1}{D(p)}=p^2+m^2+g^2\pi_1(p^2,m^2)
\end{equation}
at lowest order, where the value of $\pi_1(p^2,m^2)$ can be
extracted from \eqref{pi1}, we can solve for $m^2$ in an iterative
way as follows
\begin{equation}\label{prop4}
    m^2=\xi(p)-g^2\pi_1(p^2,\xi)\,,
\end{equation}
where we put $\xi(p)=\frac{1}{D(p)}-p^2$ for notational simplicity.
Assuming that $\xi_\ast(p)$ obeys
\begin{equation}\label{prop5}
    \xi_\ast(p)-g^2\pi_1(p^2,\xi_\ast(p))=0\,,
\end{equation}
we arrive at the following 1-loop estimate for the $3D$ gluon
propagator
\begin{equation}\label{prop6}
   D(p)=\frac{1}{p^2+\xi_\ast(p)}\,.
\end{equation}
We plotted $D(p)$ as a function of $p$ in Figure 5.

Since we are still using a series expansion, one might wonder if we
have any control over this expansion? We recall that the formal
counting of the orders in the expansion is done by $g^2$, which
unfortunately carries a dimension, hence $g^2$ is not really
suitable as expansion parameter. However, taking a look at the
inverted series \eqref{prop5}, we notice that we can say that
$\frac{g^2}{\sqrt{\xi_*}}$ will emerge as a natural
\emph{dimensionless} expansion parameter at each order\footnote{This
amounts to a dynamically realized version of the naively expected
$\frac{g^2}{m}$ in the presence of a regulating mass parameter
$m$.}. In addition, in a $D$-dimensional space time and in the
absence of quarks ($N_f=0$), the coupling constant is always
accompanied by an additional suppressing factor
$\frac{N}{(4\pi)^{D/2}}$ due to the loop integrations. In Figure 6,
we have therefore plotted the quantity
$y(p)=\frac{g^2N}{(4\pi)^{3/2}\sqrt{\xi_*(p)}}$ in function of the
momentum $p$. We recognize this is a rude way of estimating the
acceptability of a perturbative approach, but at least we can be
satisfied that $y$ is sufficiently small if we do not come too close
to zero momentum.

\begin{figure}[t]
\begin{center}
\includegraphics[width=8cm]{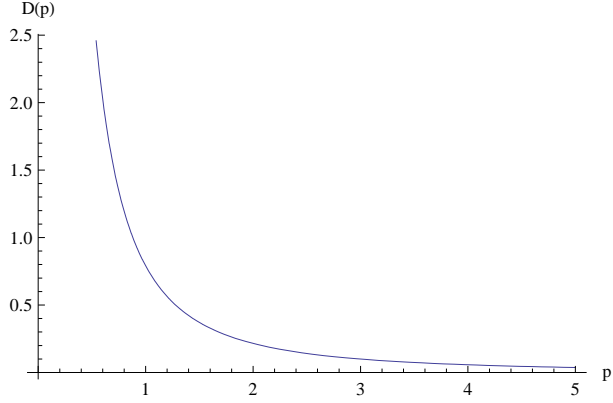}
\caption{\emph{The gluon propagator $D(p)$, in units $g^2=1$, for $N=3$.}}
\end{center}
\end{figure}

\begin{figure}[t]
\begin{center}
\includegraphics[width=8cm]{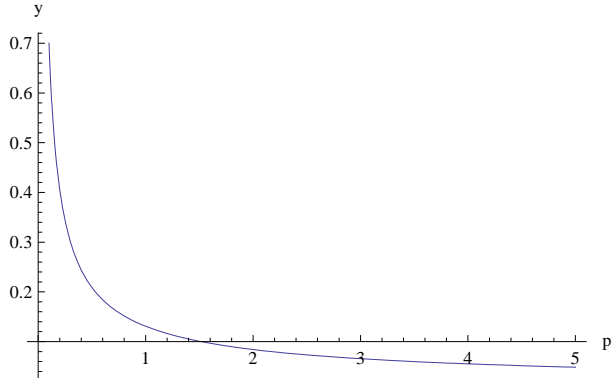} \caption{\emph{The expansion parameter
$y$, for $N=3$.}}
\end{center}
\end{figure}

\subsection{A few extra words on $A^2_{\min}$ beyond the Landau gauge}
We recall that $A^2$ is not gauge invariant, but in the Landau gauge
it equals $A^2_{\min}$, and we motivated that the latter gauge
invariant quantity is power counting renormalizable in $3D$,
assuring that we in principle considered a gauge invariant
regularization. We draw attention to the fact that the ghost in the
Landau gauge is still massless, and might introduce additional
infrared instabilities\footnote{These were not observed during the
two-loop calculations reported in \cite{Dudal:2006ip}, despite the
presence of massless ghosts in the Landau gauge.}. In order to
overcome these, one might consider using the Curci-Ferrari gauge, a
generalization of the Landau gauge, in which case the ghosts also
attain a mass \cite{Dudal:2004ch,Dudal:2006ip}. At the end of such a
calculation, the limit to the Landau gauge can be considered.
However, since the ghost mass is explicitly gauge parameter
dependent, it will not enter gauge invariant quantities, and as
such, the Landau gauge could be immediately used to calculate gauge
invariant quantities without risking extra infrared divergences.

The question remains what to do if we would like to calculate e.g.
the gluon propagator $D(p)$ in the linear covariant gauges with
$A^2_{\min}$? Infrared singularities coming from the
$\frac{1}{\p^2}$'s in \eqref{serie2} can be avoided by using
\begin{eqnarray}
S_m&=&m^2\mbox{Tr}\int \d^{3}x\left( F_{\mu \nu
}\frac{1}{D^{2}+\rho^2}F_{\mu \nu
}+2ig\frac{1}{D^{2}+\rho^2}F_{\lambda \mu }\left[
\frac{1}{D^{2}+\rho^2}D_{\kappa }F_{\kappa \lambda
},\frac{1}{D^{2}+\rho^2}D_{\nu }F_{\nu \mu }\right] \right.
\nonumber \\
&&-2ig\left. \frac{1}{D^{2}+\rho^2}F_{\lambda \mu }\left[
\frac{1}{D^{2}+\rho^2}D_{\kappa }F_{\kappa \nu
},\frac{1}{D^{2}+\rho^2}D_{\nu }F_{\lambda \mu }\right] \right)
+\mathcal{O}(F^4)  \label{serie2bis}
\end{eqnarray}
instead of \eqref{serie2}. We introduced an extra mass parameter
$\rho^2$, but neither the gauge invariance nor good UV behaviour are
compromised by this\footnote{Notice that a similar kind of
modification would not be possible using the series \eqref{min1}, as
this would spoil the gauge invariance.}. The potentially dangerous
$\frac{1}{\p^2}$ will get replaced by the IR safe
$\frac{1}{\p^2+\rho^2}$. As such, we will get a well defined
$D(p,m^2,\rho^2)$. Once this is done, one can check whether the
limit $\rho^2\to0$ exists or not. In the Landau gauge, this should
be the case as already explained before, in which case we are back
to $A^2_{\min}$. If not, one can perform a first inversion with
respect to $\rho^2$ to find a sensible $D(p,m^2)$, starting from
which a second inversion can be done to ensure $m^2=0$.

\section{Discussion}
The purpose of this paper was to illustrate that inclusion of the
operator $A^2$ in combination with the inversion method allows for a
consistent (i.e. infrared protected) perturbative expansion of
basically any quantity. However, there are other major sources of
effects beyond (regularized) perturbation theory. For example, we
can mention the existence of Gribov (gauge) copies in the Landau
gauge. Trying to restrict the integration measure in order to take
these copies into account, can have a profound influence on the
gluon propagator, analogously as in $4D$ \cite{Gribov:1977wm}. The
restriction introduces another mass scale $\gamma$ in the theory
\cite{Cucchieri:2004mf}, and we can expect that $\gamma\propto g^2$
in $3D$. A standard Gribov-like propagator looks like
$\frac{p^2}{p^4+\gamma^4}$, clearly having 2 complex poles at
$p^2=\pm i \gamma^2$.  The Gribov restriction is important as a
theoretical tool to find a violation of positivity in the gluon
propagator, indicative of confinement \cite{Cucchieri:2004mf}. In
this light, dynamical gluon mass scales, complex or real, should not
be directly related to massive ``physical'' gluons, as these are
confined and hence unphysical. Therefore, a complex pole mass as
found here or in other works \cite{Jackiw:1995nf,Jackiw:1997jg} is
not necessarily a catastrophe.  We recall that $3D$ gauge theories,
although that the classical/perturbative interaction potential is
already mildly (viz. logarithmically) confining, also display
``true'' confinement trough a linear potential, see e.g.
\cite{Teper:1998te}. The origin of this piece of the potential is
not evident.

When we compare our propagator displayed in Figure 5 with the
lattice result of \cite{Cucchieri:2003di}, then it is immediately
clear that the big difference is located in the deep infrared: the
lattice results indicate a \emph{finite} gluon propagator near zero
momentum. But at larger momenta, the inversion mechanism described
in this paper which regulates the theory can give acceptable
results.

To conclude, it would be recommendable to pursue e.g. an analytical
study in $3D$ based on \cite{Dudal:2007cw} taking into account the
existence of Gribov copies, and find out whether a more
qualititative agreement with the available lattice data can be found
also in $3D$ \cite{Cucchieri:2004mf,cucchieri}. At the same time, it
can be investigated whether the Gribov scale $\gamma$ would serve as
a natural IR regulator.

\section*{Acknowledgments}
The Conselho Nacional de Desenvolvimento Cient\'{\i}fico e
Tecnol\'{o}gico (CNPq-Brazil) and the SR2-UERJ are gratefully
acknowledged for financial support. D.~Dudal is a Postdoctoral
Fellow and N.~Vandersickel a PhD Fellow of the Research Foundation -
Flanders (FWO). D.~Dudal and N.~Vandersickel would like to thank the warm
hospitality at the UERJ and the University of Liverpool where parts of this work were done.

\appendix
\section{The Ward identity for the gluon self energy}
We start from the complete classical action
\begin{eqnarray}
\Sigma&=&S_{\mathrm{YM}}+S_{\mathrm{Landau}}+S_{\mathrm{ext}}+s\int \d^3\!x\,\biggl(\frac{1}{2}\tau A^{a}_{\mu}A^{a}_{\mu}\biggr)\nonumber\\
&=&\int \d^3\!x\,\biggl(\frac{1}{4}%
F^{a}_{\mu\nu}F^{a}_{\mu\nu}+b^{a}\partial_{\mu}A^{a}_{\mu}+\overline{c}%
^{a}\partial_{\mu}D_{\mu}^{ab}c^{b}+\frac{1}{2}JA^{a}_{\mu}A^{a}_{\mu}+\tau
A^{a}_{\mu}\partial_{\mu}c^{a}-\Omega^{a}_{\mu}D_{\mu}^{ab}c^{b}+%
\frac{g}{2}f^{abc}L^{a}c^{b}c^{c}\biggr)\;, \label{YM+GF+LCO}
\end{eqnarray}
supplemented with the necessary extra (external) source terms, e.g.
$J$ which is used to couple the operator $A^2$ to the theory.
Moreover, $s$ denotes the usual BRST symmetry generator,
\begin{eqnarray}
sA_{\mu }^{a}=-\left( D_{\mu }c\right) ^{a}\,,\qquad  sc^{a}
=\frac{1}{2}gf^{abc}c^{b}c^{c}\,,\qquad   s\overline{c}^{a}
=b^{a}\,,\qquad sb^{a}=0\,,
\end{eqnarray}
extended to the sources by means of
\begin{eqnarray}
s\tau=J \,,\qquad sJ=0\,,\qquad s\Omega^{a}_{\mu}=0\,,\qquad
sL^{a}=0.
\end{eqnarray}
The corresponding Slavnov-Taylor identity reads
\begin{equation}
\mathcal{S}(\Sigma )=\int \d^{3}\!x\,\biggl(\frac{\delta \Sigma
}{\delta \Omega _{\mu }^{a}}\frac{\delta \Sigma }{\delta A_{\mu
}^{a}}+\frac{\delta \Sigma }{\delta L^{a}}\frac{\delta \Sigma
}{\delta c^{a}}+b^{a}\frac{\delta \Sigma }{\delta
\overline{c}^{a}}+J\frac{\delta \Sigma }{\delta \tau }\biggr)=0\;.
\label{ST}
\end{equation}
The Ward identity for the vacuum polarization can be derived from
this Slavnov-Taylor identity, suitably extended to the quantum
level. At the one loop level, one has
\begin{equation}
\Gamma =\Sigma +\hbar \Gamma ^{1}\;,  \label{b13}
\end{equation}
so that the 1-loop Slavnov-Taylor identity becomes
\begin{eqnarray}
\int \d^{3}x\left( \frac{\delta \Gamma ^{1}}{\delta \Omega _{\mu }^{a}}\frac{%
\delta \Sigma }{\delta A_{\mu }^{a}}+\frac{\delta \Sigma }{\delta
\Omega _{\mu }^{a}}\frac{\delta \Gamma ^{1}}{\delta A_{\mu
}^{a}}+\frac{\delta \Gamma ^{1}}{\delta L^{a}}\frac{\delta \Sigma
}{\delta c^{a}}+\frac{\delta \Sigma }{\delta L^{a}}\frac{\delta
\Gamma ^{1}}{\delta c^{a}}+b^a\frac{\delta \Gamma ^{1}}{\delta
\overline{c}^a }+J\frac{\delta \Gamma ^{1}}{\delta \tau }\right)
=0\,.\nonumber\\\label{stid}
\end{eqnarray}
Using\footnote{$\left[ \mathcal{O}\cdot \Gamma \right] $ denotes the
generator of the 1PI Green functions with the insertion of the
composite operator $\mathcal{O}$.}
\begin{eqnarray}
\frac{\delta \Gamma ^{1}}{\delta \Omega _{\mu }^{a}} &=&\left[
-\left( D_{\mu }^{ab}c^{b}\right) \cdot \Gamma \right] ^{1}\;,\quad
\frac{\delta \Gamma ^{1}}{\delta L^{a}} =\left[ \left( \frac{g}{2}%
f^{abc}c^{b}c^{c}\;\right) \cdot \Gamma \right] ^{1}\;, \quad
\frac{\delta \Gamma ^{1}}{\delta \tau} =\left[\left(A_\mu^a\p_\mu
c^a\right) \;\cdot \Gamma \right] ^{1}\,,  \label{b14}
\end{eqnarray}
we derive
\begin{eqnarray}
&&\int \d^{3}x\left( \left[ -\left( D_{\mu }^{ab}c^{b}\right) \cdot
\Gamma \right] ^{1}\frac{\delta \Sigma }{\delta A_{\mu }^{a}}-\left(
D_{\mu }^{ab}c^{b}\right) \frac{\delta \Gamma ^{1}}{\delta A_{\mu
}^{a}}+\left[ \left( \frac{g}{2}f^{abc}c^{b}c^{c}\;\right) \cdot
\Gamma \right]
^{1}\frac{\delta \Sigma }{\delta c^{a}}\right.  \nonumber \\
&+&\left.\left( \frac{g}{2}f^{abc}c^{b}c^{c}\;\right) \frac{%
\delta \Gamma ^{1}}{\delta c^{a}}+ J\left[\left(A_\mu^a\p_\mu
c^a\right) \;\cdot \Gamma \right] ^{1}+ b^{a}\frac{\delta \Gamma
^{1}}{\delta \overline{c}^{a}}\right)=0\,.  \label{b15}
\end{eqnarray}
Applying the test operator $ \frac{\delta ^{2}}{\delta
c^{a}(x)\delta A_{\nu }^{b}(y)}$ and setting all fields and sources
equal to zero at the end of the operation, except for $J=m^2$, one
obtains the Ward identity for the vacuum polarization
$\pi_{\mu\nu}^{ab}$
\begin{equation}
\p_\mu \pi_{\mu\nu}^{ab}(x,y)\equiv\partial _{\mu }^{x}\frac{\delta
^{2}\Gamma ^{1}}{\delta A_{\mu }^{a}(x)\delta
A_{v}^{b}(y)}=-m^2\left( \frac{\delta ^{2}\left[ \displaystyle\int
\d^{3}z\left(A_\mu^a\p_\mu c^a\right)
_{z}\;\cdot \Gamma \right] ^{1}}{\delta c^{a}(x)\delta A_{\nu }^{b}(y)}%
\right)+ m^2\left( \frac{\delta \left[
\displaystyle\left(D_\nu^{bd}c^d\right)
_{y}\;\cdot \Gamma \right] ^{1}}{\delta c^{a}(x)}%
\right)\,.  \label{b17}
\end{equation}
The first term is trivial, by using partial integration and the fact
that we are considering the Landau gauge $\p A=0$. Hence, the vacuum
polarization in the massive Landau case is \emph{not} transverse,
but rather subject to the following Ward identity at one loop
\begin{equation}
\p_\mu \pi_{\mu\nu}^{ab}(x,y)=m^2\left( \frac{\delta \left[
\displaystyle\left(D_\nu^{bd}c^d\right)
_{y}\;\cdot \Gamma \right] ^{1}}{\delta c^{a}(x)}%
\right)\,.  \label{b17b}
\end{equation}

\end{document}